\documentclass[]{article}
\usepackage[utf8]{inputenc}
\usepackage{graphicx}
\usepackage{longtable}
\usepackage{color}
\usepackage{amsmath}
\usepackage{xcolor}
\usepackage{hyperref}
\usepackage{authblk}
\usepackage{subcaption}
\usepackage{geometry}
\usepackage{multicol}
\usepackage{changepage}
\usepackage{caption}
\usepackage{url}

\newcommand\review[1]{{\color{black}#1}}

\title{Cities beyond proximity}

\author[1]{Dan Hill}
\author[2]{Matteo Bruno}
\author[2,3]{Hygor P. M. Melo}
\author[4,5]{Yuichiro Takeuchi}
\author[2,6,7]{Vittorio Loreto\thanks{Corresponding author: \texttt{vittorio.loreto@sony.com}}}

\affil[1]{Faculty of Architecture, Building, and Planning, University of Melbourne, Australia}
\affil[2]{Sony CSL - Rome, Joint Initiative CREF-SONY, CREF, Via Panisperna 89/A, 00184, Rome, Italy}
\affil[3]{IFCE, Av. Des. Armando de Sales Louzada, Acaraú, Brazil}
\affil[4]{Sony CSL - Kyoto, 13-1 Hontorocho, Shimogyo Ward, 6008086, Kyoto, Japan}
\affil[5]{Wikitopia Institute, 214-1 Nakakanabutsucho, Shimogyo Ward, 6008332, Kyoto, Japan}
\affil[6]{Sapienza Univ. of Rome, Physics Dept, Piazzale A. Moro, 2, 00185, Rome, Italy}
\affil[7]{Complexity Science Hub, Josefst\"{a}dter Strasse 3, A 1080, Vienna, Austria}

\begin{document}

\maketitle

\begin{abstract}
The concept of `proximity-based cities' has gained attention as a new urban organizational model. Most prominently, the 15-minute city contends that cities can function more effectively, equitably and sustainably if essential, everyday services and key amenities are within a 15-minute walk or cycle. However, focusing solely on travel time risks overlooking disparities in service quality, as the proximity paradigm tends to emphasize the mere presence of an element in a location rather than bringing up more complex questions of identity, diversity, quality, value or relationships. Transitioning to value-based cities by considering more than just proximity can enhance local identity, resilience and urban democracy. Fostering bottom–up initiatives can create a culture of local care and value, while predominantly top–down governing strategies can lead to large inequalities. Balancing these approaches can maximize resilience, health and sustainability. This equilibrium has the potential to accompany sustainable growth, by encouraging the creation of innovative urban solutions and reducing inequalities.
\end{abstract}

\maketitle

\section{Introduction: stating the problem and state of the art}

The ‘proximity-based cities’ theme has attracted much attention in the framework of the quest for new paradigms for urban organisation, regeneration and renewal~\cite{batty2013new,batty2024computable}. Most prominently, the 15-minute city~\cite{moreno2021introducing,moreno2020ville} contends that cities can function in more effective, equitable, and environmentally-conscious ways if essential, everyday services and key amenities are within a proximity of ~15 minutes by active forms of transport, such as walking, cycling, or rolling. 

This notion belongs to a broader framework known as chrono-urbanism, where the most crucial dimension for planning is time~\cite{moreno2021introducing}. This model proposes a utopian vision where the decentralisation of activities and opportunities engenders a form of polycentric urbanism, in which multiple centres of activity reduce the need for extensive travel and promote sustainability. This proximity-based thinking powerfully presents alternatives to previous existing patterns of urban organisation, which have tended to contribute to, if not drive, forms of urban sprawl, a key factor in today’s systemic challenges of climate breakdown, social fragmentation and poor public health~\cite{brown2009planning, gossling2020cities, brown2003redesigning}. 

The 15-minute city model aligns with historical urban planning ideologies, drawing from Howard’s Garden City, which emphasises the significance of accessibility and local centrality~\cite{howard2003garden}. The enduring influences of Jacobs’ urban vitality, Calthorpe and Fulton’s New Urbanism, Hägerstrand’s Time Geography, and Gehl’s human-scale design are woven into the current narrative advocating for proximity in urban planning~\cite{jacobs1989death,calthorpe2001regional, hagerstrand1976geography, hagerstrand1981space,gehl2013cities}.

\begin{figure}[h]
\centering
\includegraphics[width=\columnwidth]{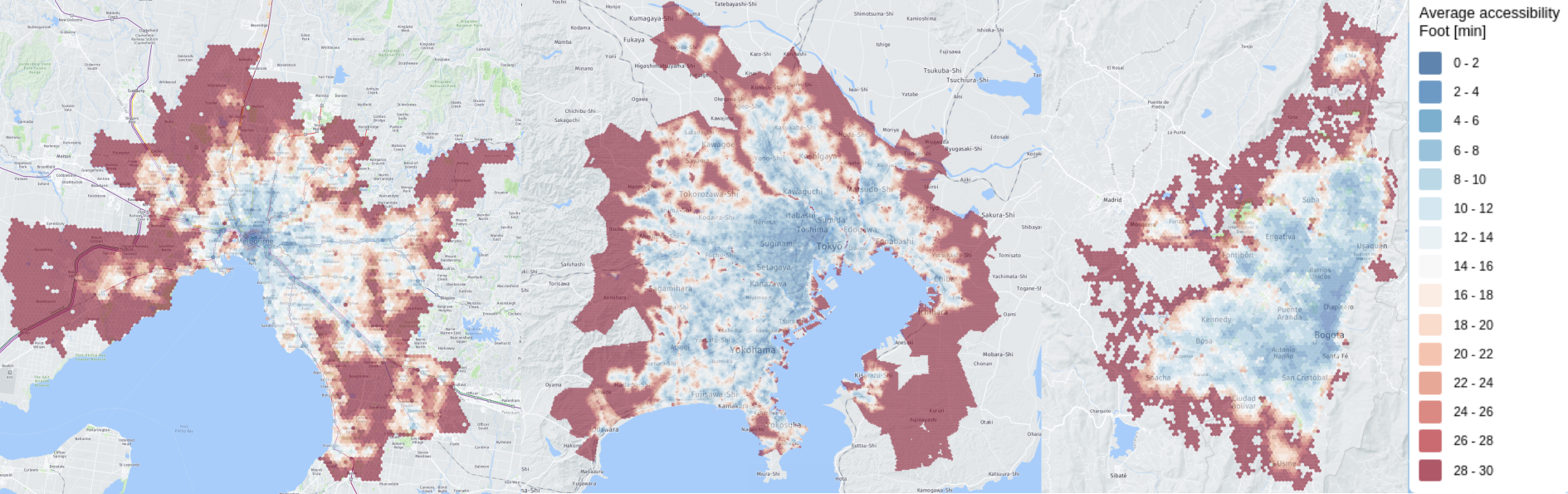}
\caption{\textbf{A visualisation of the 15-minute proximity times for Melbourne, Tokyo and Bogot\'a.} Blue areas are the ones that mostly adhere to the 15-minute paradigm, while red areas have a scarcer availability of services. White areas are at the threshold, with a 15-minute average time of accessibility. A radial pattern is usually noticeable, with central areas disposing of better availability of services.}
\label{fig:15min_cities}
\end{figure}

Before we proceed any further, let us present an assessment of the closeness of current cities to the ideal of 15-minuteness. Some of us proposed an online platform (\href{whatif.sonycsl.it/15mincity}{whatif.sonycsl.it/15mincity}) to access and visualise accessibility scores for virtually all cities worldwide. The platform allows us to visualise the accessibility times to resources and services. Fig.~\ref{fig:15min_cities} reports examples of the visualisation of the 15-minuteness of three cities, Melbourne, Tokyo and Bogot\'a. Each hexagon in each city is coloured according to the Proximity Time, i.e., the time to access the services in the basket of the 15-minute city. The colour code is such that blueish (reddish) colours correspond to areas whose accessibility time of the services is below (above) 15 minutes.

The analysis performed on many cities worldwide reveals strong heterogeneity of accessibility both within and across different cities, with a significant role played by local population densities. The heterogeneity of accessibility within cities is one of the sources of inequality.

Despite the intuitive appeal of proximity-based urbanism, several critical limitations warrant attention:
\begin{description}

\item[Limited agency] Several critical urban issues, such as gentrification and social equity, accessibility issues beyond mere proximity, lack of affordable housing, meaningful participation in shared decision-making, cultural diversity, regional economic patterns, etc., lie outside a proximity-based scope and cannot be addressed, at least directly.

\item[Limited granularity] Proximity-based analyses are unable to provide meaningful insights for very high-density environments (e.g., they fail to describe the quality-of-life differences among central Tokyo neighbourhoods adequately), due to a lack of facility with the specific qualities of particular spaces, services, amenities and environments. 

\item[Limited applicability] The paradigm of a proximity-based city is challenging to implement in existing rural areas or lower-density suburban neighbourhoods. In these environments, a proximity-based organisation generally requires layers of densification, which may be unfeasible due to socioeconomic and/or political constraints.

\end{description}

Such limitations stem from the fact that analyses predominantly focusing on proximity are, in essence, emphasising a few relatively easily quantifiable aspects of urban life (e.g. \textit{Is there a swimming pool within 15 minutes? Is there a train station within 15 minutes?}) whilst ignoring others (e.g. \textit{'Is the swimming pool welcoming to diverse swimmers? Does the train station have a regular, reliable service?}). These factors clearly influence whether these services and spaces are truly open, available, and likely to be used well beyond their mere existence 'within range'. 

\review{The ideal of proximity is further challenged by the rise of door-to-door services. Although these services provide convenience, they can undermine local engagement, local economies, and diminish the vibrancy of neighbourhoods. Online shopping and food deliveries have increasingly captured significant market share from traditional proximity services as clothing shops, hardware stores, restaurants, and even supermarkets transition to online delivery systems. Yet the proximity-based approach also allows urban planners and policymakers to orchestrate production, consumption and logistics services across a city. Cities like Utrecht, with their integrated area-based freight Hub Utrecht Oost, indicate more sustainable ways of coordinating deliveries. Equally, the x-minute city model—when expanded from 1-minute to 15-minute to 60-minute and beyond—provides a framework for policymakers who might incentivise local production and sustainable delivery. Clearly, urban planners and policymakers must consider this paradigm shift when designing neighbourhood density, shaping local economic models, incentivising local production or organising sustainable service offerings, in order to mitigate the risk to local economies and environments.}

Moreover, focusing only on proximity neglects an essential element of a city's so-called Points of Interest (POIs), namely their hierarchical nature. We shall come back to this point later on more extensively. It is enough to mention that POIs feature a hierarchical structure in terms of their importance and uniqueness\cite{guzman2024proximity}. While one could have a grocery store at virtually every corner, the same cannot be true for large facilities, e.g., a prominent theatre or a stadium, or iconic landmarks, e.g., the Coliseum in Rome or the Empire State Building in New York City. Given this unavoidable hierarchical nature, focusing only on proximity might be misleading because it would imply that every neighbourhood should have all types of amenities. We know this is not possible. 
From this perspective, the 15-minute city overly simplifies the needs and desires of people living in urban areas: access to multiple, non-local opportunities\cite{abbiasov2022quantified}. A broader concept of \textit{opportunity city} could be more suitable to address this neglected behavioural aspect. 
 
Trying to address some of the concerns mentioned above, a body of literature has emerged to fill the theoretical void surrounding the concept of the proximity city. Some studies have sought to map accessibility and quantify inequalities by geolocating amenities and calculating service accessibility times across urban landscapes~\cite{15minwebsite,vale2023accessibility,nicoletti2023disadvantaged}. This approach can determine which cities are closer to the ideal of a 15-minute city and which are not. Of course, this quantification helps photograph the inequality of accessibility in cities and urban areas, yet they often fail to capture the full spectrum of urban life and its diversity. The quantification of proximity is realised with different methods~\cite{megahed2024reconceptualizing,lima2023quest}. Still, they are primarily based on practical space/time counts of services reachable by walking, cycling, or equivalent. Thus, they fail to be comprehensive, inclusive, and representative of the qualities and values of the actual urban fabric and the diversity of living patterns.

Critiques and analyses highlight the complexity of implementing this model universally. An extensive numerical analysis demonstrated that the overall population density for some cities can be so low that a 15-minute city is unfeasible~\cite{bruno15min2024}. On the other hand, a critical examination suggests that individual perceptions of what constitutes 'proximity' vary significantly, influenced by personal and cultural factors that affect how distance and convenience are perceived~\cite{guzman2024proximity}. This variability challenges the application of a standardised 15-minute threshold across diverse populations and geographic contexts~\cite{staricco202215,logan2022x}.

An empirical study using human mobility data further quantifies the challenge of realising a 15-minute city. Analysis of urban mobility patterns in the USA reveals significant deviations from the 15-minute ideal -- even when services are accessible within 15 minutes, people may still choose to use services located further away creating large spatial and social-economical disparities~\cite{abbiasov202415}. Such findings underscore the gap between the theoretical model and its practical implementation in current urban landscapes.

Focusing on specific case studies, research conducted in Barcelona illustrates how varying urban layouts influence the practicality of the 15-minute city. The study reveals that while certain districts align well with the model, others do not, reflecting broader issues of urban inequality based on underlying and deep-seated socio-economic disparities and the need for tailored approaches in urban planning~\cite{graells2021city}. The precise details of an urban environment's topography also directly affect the perception and reality of such 15-minute city ideas. Neighbourhood access cannot be accessed by a 'crows-fly' model of proximity measurement, given that geographic or topographic conditions often exacerbate socio-economic disparities. In Lima, Peru, low-income habitation is usually located on hillsides or sloping land, which reveals differentials in what distance '15-minutes of walking' approximate to~\cite{silva2023capturing}. Even in pedestrian-friendly cities like Barcelona, considering various pedestrian mobility profiles is essential to assess the feasibility of a 15-minute city model, as demonstrated in~\cite{rhoads2023inclusive}. Aligning a detailed understanding of built form and topography associated with socio-economic analysis further reinforces the need for more nuanced measures.      

\begin{figure}
    \centering
    \includegraphics[width=1\linewidth]{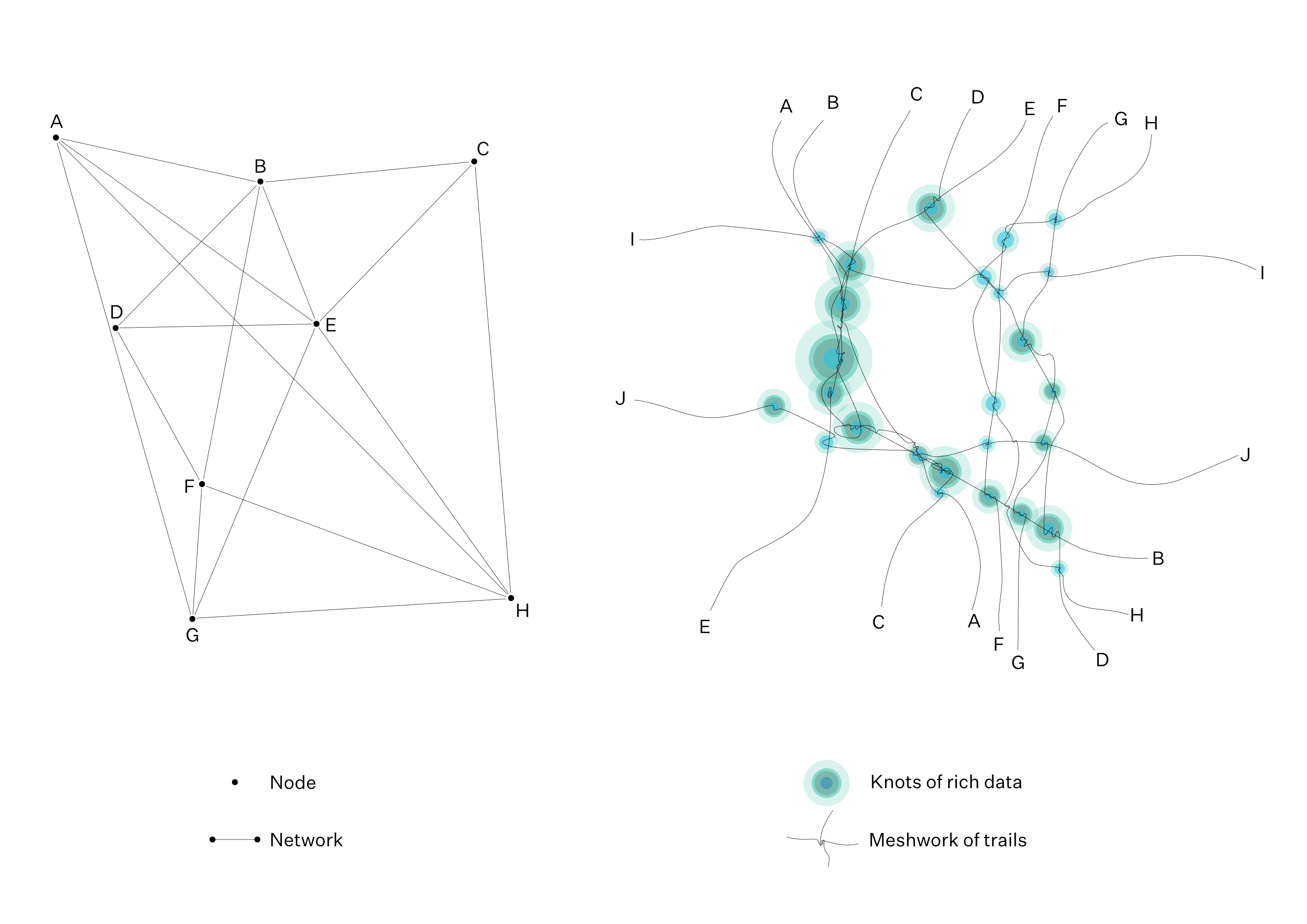}
    \caption{Conventional proximity-based analysis, with simplistic data about place and path (left); Qualitative rich data about tangled knots in a meshwork of trails, after Ingold (right)}
    \label{fig:2}
\end{figure}

These theories, of course, rely on movement across and between those proximities. Regarding such movement, the anthropologist Tim Ingold has drawn subtle distinctions between "the walk" and "the assembly"~\cite{ingold2007lines}. Proximity cities exemplify what Ingold describes as the shift to seeing movement as "a succession of points or dots ... (which has) transformed our understanding of place: once a knot tied from multiple and interlaced strands of movement and growth, it now figures as a node in a static network of connectors." (See Figure~\ref{fig:2}). In this, Ingold suggests we have lost a more diversified and complex sense of movement. He differentiates indigenous knowledge systems and their way of seeing from those of "the colonial project of occupation", comparing Inuit hunters and the colonial British naval high command, and thus the difference between "wayfaring" and mere "transport". 

As proximity cities reduce this rich complexity about both place and movement down to a simplistic network of nodes, we see a narrowing not only of experience but also of space, place, and the way that these things are knotted together. In its place, Ingold suggests that through "metaphorical extension to the realms of modern transport and communications, and especially information technology, the meaning of 'the net' has changed". This has reduced the rich data of "trails" to a shallower reading of "routes”, according to Ingold.

The qualitative perspective of tangled knots in trails have been lacking thus far in a proximity cities framework. Ingold writes, "I suggest that to understand how people do not just occupy but inhabit the environments in which they dwell, we might do better to revert from the paradigm of the assembly to that of the walk.” (Chap. "Up, Across and Along" in~\cite{ingold2007lines}).

Another perspective reinforces the need for a richer understanding of access. A comparative analysis across European cities highlights substantial accessibility inequalities~\cite{vale2023accessibility}. This study uses pedestrian accessibility metrics to demonstrate that not all European cities conform to the 15-minute model, and significant within-city and between-city inequalities persist. In Ref.~\cite{bruno15min2024}, the accessibility times to resources and services have also been measured, and the outcomes reveal strong heterogeneity of accessibility within and across cities, with a significant role played by local population densities. Since the heterogeneity of accessibility within cities is one of the sources of inequality, it was also simulated how much one could 'heal' inequity by redistributing the same number of resources and services or by allowing for virtually infinite resources. The general conclusion was that proximity-based paradigms must be generalised to work on a wide range of local population densities and socio-economic and cultural factors. These findings advocate for more nuanced measures, considering the variety of urban opportunities and their equitable distribution.

As discussed, the current literature demonstrated that while the 15-minute city model offers a promising framework for sustainable urban development, its success hinges on acknowledging and addressing the diverse needs, perceptions, and spatial, cultural and political contexts of urban populations. These studies collectively emphasise the importance of implementing adaptable, context-aware planning strategies that recognise cities' unique spatial and social fabrics.

The adoption of the 15-minute city model presents both opportunities and challenges. Urban planners, designers, and policymakers must consider accessibility as a function of individual, community, and spatial diversity to ensure that the transition towards sustainable urban environments is inclusive and equitable. Future research should continue exploring how adaptive urban planning and design can accommodate these complexities, ensuring that the 15-minute city model benefits all urban residents equally.

To overcome these problems, we propose expanding the proximity-based paradigm by exploring additional layers of \textit{value-based analysis}. More precisely, we assign additional parameters to POIs (Points of Interest, i.e., urban facilities and amenities whose geographical locations constitute the central focus of proximity-based analyses), which might begin to reflect their qualitative properties. This expanded paradigm must consider not only each resident's proximity to the individual POIs, but also the quality of interactions offered by the POIs and the 'trails' that thread them together, and thus begin to assess and evaluate their contributions to neighbourhood vitality, inclusiveness, and identity (See Figure~\ref{fig:2}). 

\section{The role of accessibility and participation for citizens' care}

\subsection{Care as an Urban Concept}

At the heart of our approach is the notion of "care", which we define as an overarching parameter encapsulating various aspects of urban POIs. In practice, care may be represented not necessarily by a single scalar or vector parameter but by a complex web of parameters, also including algorithmic functions, machine learning models, etc. Care encodes varied information, such as the qualities of service or interactions offered by the POI, the affordability, accessibility, diversity, or openness of the POI, the capacity for citizens to participate in the POI's operation,  maintenance, and ownership, etc. As is the case with proximity, there exists a personal and phenomenological dimension to care — that is, the degree of care that one may ascribe to a POI will depend on that person's specific profile, identity and socio-economic context. For example, one may struggle to \textit{care for} a POI that excludes them based on factors such as occupation, identity, or socio-economic status.

Note that we include the capacity for citizen participation as a key element of care. Thus, according to our definition, a community garden will have the opportunity or latent potential—the disposition, after Easterling~\cite{easterling2014extrastatecraft}—for a higher degree of care and a greater engagement level than a private garden. The latter, in contrast, prohibits citizens from taking active roles in its maintenance. Both types are gardens and will tend to be classified as such in mapping systems. Beyond their ownership status, which may be denoted in mapping systems, these notional gardens may be similar in every other feature—size, soil quality, solar aspect, diversity of planting, adjacencies, etc.—and located within the same range of people. Yet this assemblage of distinct attributes, combined with the gardens' ownership and access status, significantly shapes the possibilities for open participation by a diversity of people. The legal status of the garden in terms of ownership—public or private—is only one aspect of whether a garden feels open and approachable: in other words, capable of supporting participation.

This line of thought is informed by previous applied research projects in Sweden, developing mission-oriented innovation projects around mobility and public space~\cite{hill2022designing}. These interventions revealed how public spaces, like city streets, are generally defined mainly by distanced and abstracted professional and bureaucratic processes and instead allow residents to 'design their street'. They revealed that, given the choice, residents and users generated a greater diversity of ideas and uses for their streets as public spaces, not simply a traffic corridor or a parking space. Described as a '1-minute City' process, this was a deliberate subversion of proximity-based thinking to foreground participative approaches to city-making rather than municipal. This argument positioned the 15-minute city, for example, as a primarily municipal process of organising core urban amenities on behalf of citizens (i.e., the location of swimming pools, libraries, healthcare centres, public transport, activity centres, and so on). The 1-minute City deliberately skewed decision-making cultures to the highly immediate, intimate range of \textit{'the street outside your front door',} which could be a participative process negotiated and organised directly with neighbours and frequent users. Similarly, these 1-minute City environments could also be cared for and maintained by neighbours and frequent users. In the first instance, six-year-old Stockholm schoolchildren 'designed their street' outside their schoolyards, facilitated by professional designers. This highly participative approach not only led to increased utilisation of the street (by as much as 400 per cent) but also a greater diversity of users and uses on the street, with a distinct shift of urban amenities from car-based infrastructures like on-street parking towards social spaces, seating, shared bicycle and scooter parking, greenery and so on~\cite{hill2022designing}.

As we will describe, this suggests the idea of a spectrum of participation and representation within urban environments, from scales loosely defined by 1-minute to 15-minute and beyond, which clearly moves beyond mere proximity, and towards an understanding of the qualitative and particular forms of participation, openness and care.

\subsection{Evaluating Care}

As mentioned earlier, our definition of care is emerging and necessarily incomplete, yet already encompasses a wide range of properties that may be ascribed to POIs, enriching and extending our understanding of what a POI might mean within the context of a neighbourhood or city. 

This renewed emphasis on the key \textit{qualitative} differences in different aspects of proximity cities—mobility, learning, healthcare, culture, employment, recreation, etc.—is pursued with the intent to understand these varying qualities better, producing a more holistic, resilient, and adaptive urbanism; in short, an ‘open city’~\cite{sennett_2017}, by better how to represent, and value, \textit{forms of participative care} in the city~\cite{manzini2022livable}, introducing this crucial aspect of urbanism into the discourse of proximity cities. This inclusion may enable an enriched understanding, overlaying currently simplistic proximity-based approaches with a more nuanced spectrum of care, moving between participative and representative relationships within urban systems.

Building on the well-understood values of ‘compact cities’ and the spatial, social, and environmental qualities of mixed-use neighbourhoods in this sense — and understanding what qualities afford forms of participation and care — these new ‘analytical lenses’ may provide a mechanism for identifying, describing, curating, and maintaining/managing the value of distributed, open, participative systems and spaces — often at fine grain or small scale and with fluid and dynamic usage, perhaps newly-enabled by digital services. It could provide a meaningful impetus and rationale for introducing such elements into existing and new cities by overlaying the key question of urban relationships onto the proximity- and presence-based measures.

Below is an initial list of such subcategories of care that may be directly measured, referenced, or inferred relatively straightforwardly. This is presented as an initial suggested list to be subsequently tested and refined through experiments and prototypes.
\begin{table}[h!]
\begin{center}
\begin{tabular}{ |p{5.8cm}|p{6.2cm}| } 
\hline
 \textbf{Urban element lens} & \textbf{Descriptive attributes}\\ 
 \hline
 Distribution, diversity and number & Are there multiple varying instances of these elements? How many diverse mixed uses are within this element? What is the urban functional mix that this element sits within\\ 
 \hline
 Grain and porosity & What is the grain of the patterning? Is it porous or closed? What is the block size? \\ 
 \hline
 Ownership status & Is it co-owned, shared, public, or corporate, etc.? \\ 
 \hline
 Positioning and condition & Location, visibility, and accessibility on the street? \\ 
 \hline
 Legibility & Are the system’s characteristics understandable and interactive? \\ 
 \hline
 Accessibility & Is it open to more, and more diverse, types of people? \\ 
 \hline
 Adaptability and incompleteness& What is the zoning for this element? What heritage constraints, if any? Is it spatially adaptable?\\ 
 \hline
 Opening hours and frequency of service& Is it frequently accessible? \\ 
 \hline
 Capital cost & What is the risk of adapting or changing? \\ 
 \hline
 Cost of entry, or fee & Does it cost money to interact with, and how expensive is this, perhaps in relation to local economic conditions? \\ 
 \hline
\end{tabular}
\end{center}
\caption{Preliminary list of features of POIs that may be directly measured, referenced, or inferred relatively straightforwardly.}
\label{table:1}
\end{table}

Some of these properties build on previous concepts. For instance, ‘urban functional mix’ is derived from Dovey and Pafka~\cite{dovey2017functional}. Many of them tangentially relate to urban scholar Richard Sennett’s concept of the \textit{‘open city’}~\cite{sennett_2017}, with his emphasis on the spatial qualities of ‘grain, porosity, legibility, distribution and incompleteness’ (related to adaptation) that must also be seen in the context of political aspects of ownership, accessibility and ‘narrative indeterminacy’~\cite{sennett_2007}. ‘Legibility’ is also inferred from Kevin Lynch’s seminal text ‘\textit{The Image of the City’}~\cite{lynch_1964}. Other properties are more mundane and easily accessible from mapping products and websites—the cost of entry or fee, opening hours—yet clearly affect a person’s ability to fundamentally interact with a POI beyond the fact of proximity that it is ‘within range’.

Our notion of care also includes properties that are either latent or deeply subjective in nature, as listed below. Such properties often resist simple measurement or inference but may be made manifest through surveys, heterogeneous data analyses, etc.

\begin{table}[h!]
\begin{center}
\begin{tabular}{| p{5.8cm}|p{6.2cm} |} 
 \hline
 \textbf{Urban element lens} & \textbf{Descriptive attributes}\\ 
 \hline
 Service quality & Does it provide high-quality service to citizens?\\
 \hline
 General sentiment & Do people consider it generally favourable or unfavourable? Desirable or undesirable? Good or bad? Perception that is safe or unsafe?\\ 
 \hline
 Indispensability & Do people consider it an essential or expendable part of their cities or their lives? \\ 
 \hline
 Social function & Does it facilitate social interactions among diverse forms of citizens, especially of non-transactional kinds?\\ 
 \hline
 Imageability & Does it prominently feature in people’s ideas of the immediate city? Is it seen as part of the neighbourhood or simply nearby?\\
 \hline
 Maintenance & Do people consider the POI to be well-maintained, cared-for and tended?\\
 \hline
 Inclusiveness & Do people see it as accessible and inclusive to diverse peoples, or only accessible or available for particular groups?\\ 
 \hline
\end{tabular}
\end{center}
\caption{Additional properties for POIs that are either latent or deeply subjective in nature.}
\label{table:2}
\end{table}

As noted, these properties are highly qualitative and subjective—“good or bad” is clearly in the eye of the beholder—and yet many of these qualities are crucial to the perception of whether a POI is genuinely accessible, not simply proximal. Whether one can participate in a POI, whether merely using it, maintaining it, sharing it, or co-owning it, has as much to do with qualities like these as it has to do with the mere fact of its existence nearby. Incorporating such information into our analyses will require new mechanisms for data collection. In section~\ref{sec:data} we discuss the limitations of current data-based approaches.

Dedicated digital platforms, such as the Sony CSL \textit{dédédé} prototype (\href{dedede.de/en/}{dedede.de}), have been constructed to capture a looser ‘net’ of qualitative information about urban environments. These platforms are relatively open-ended in terms of input, based around photos and simple open text input, but could be subsequently analysed—via both machine vision analysis of imagery and text/metadata analysis—to infer patterns of data that may help populate some of the above properties. (Alongside such dedicated platforms, data from existing, general-purpose social media may be used as well; platforms such as X, Instagram, Facebook, TikTok, etc., can provide us with more diverse inputs as compared with dedicated urban analysis platforms, albeit with greater levels of noise.)

\begin{figure}[h]
\centering
\includegraphics[width=0.9\columnwidth]{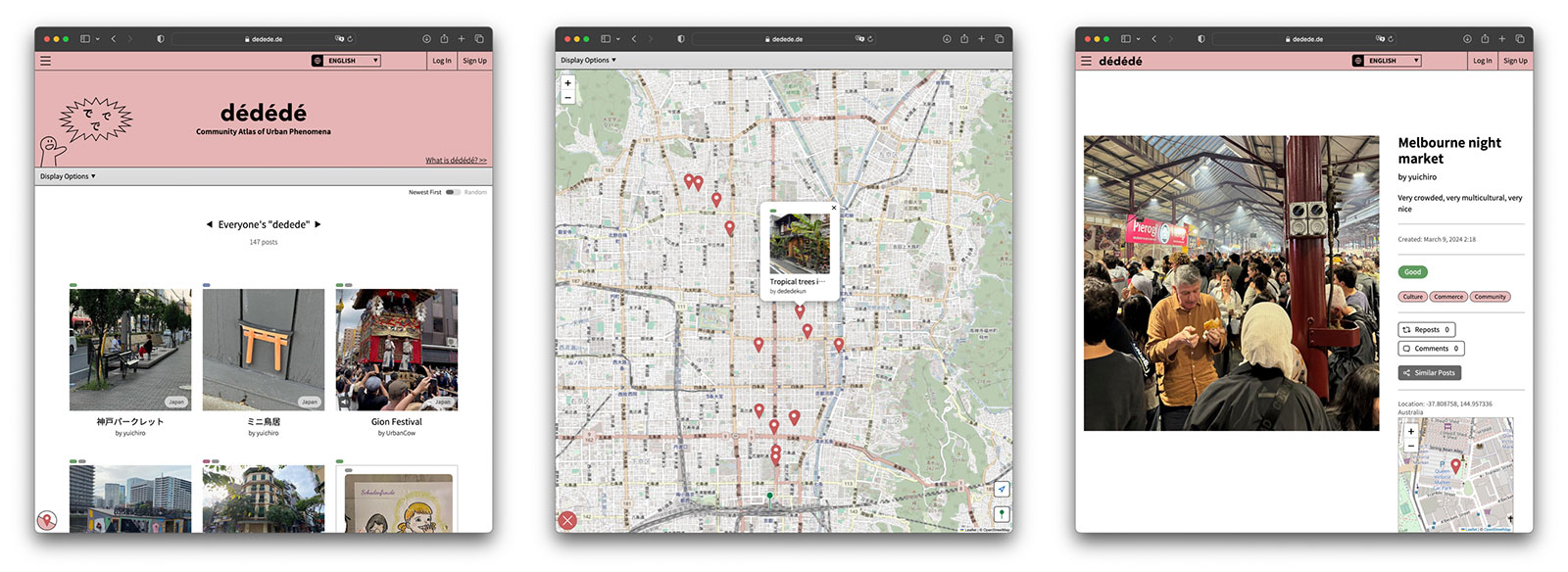}
\caption{\textbf{Screenshots of dédédé, a dedicated web-based platform for collecting qualitative opinions about cities.}}
\label{fig:dedede_ui}
\end{figure}

Based on user perception, feeling and observation, these highly qualitative aspects could complement the variables described above. If taken together, they could let emerge a far more complex, richer stew of data around POIs. Advanced data science, incorporating machine learning techniques, could work with such variable inputs. These augmented data sets and the generative aspects of their analysis could suggest these conditions for participation, belonging, identity, and possibility of care.

\subsection{Taking care, via participation or representation}

Following the line of thought of the previous paragraph, we may be able to assign insights to POIs, which reveal higher or lower degrees of care and participation, respectively. These augmented POIs, seen through this lens, allow us to understand some key differences between POIs that are absent within a traditional proximity-city framework. This point not only adds colour to an understanding of specific POIs, in some representing the vitality and diversity of urban environments, but may also provide insights as to optimal forms of governance and organisation of critical foundational points of interest, or indeed points of \textit{infrastructure}. These would include affordances or disposition for participation—and thus meaning, identity, belonging, diversity—or, conversely, which systems may be best organised through representative models, thus foregrounding resilience, equity, and long-term stability. From a system design perspective, this allows us to explore how participative decision-making cultures might be best deployed, nested within broader scales of representative decision-making, as illustrated in Figure~\ref{fig:participation_scales} below.
\begin{figure}[h]
\centering
\includegraphics[width=0.8\columnwidth]{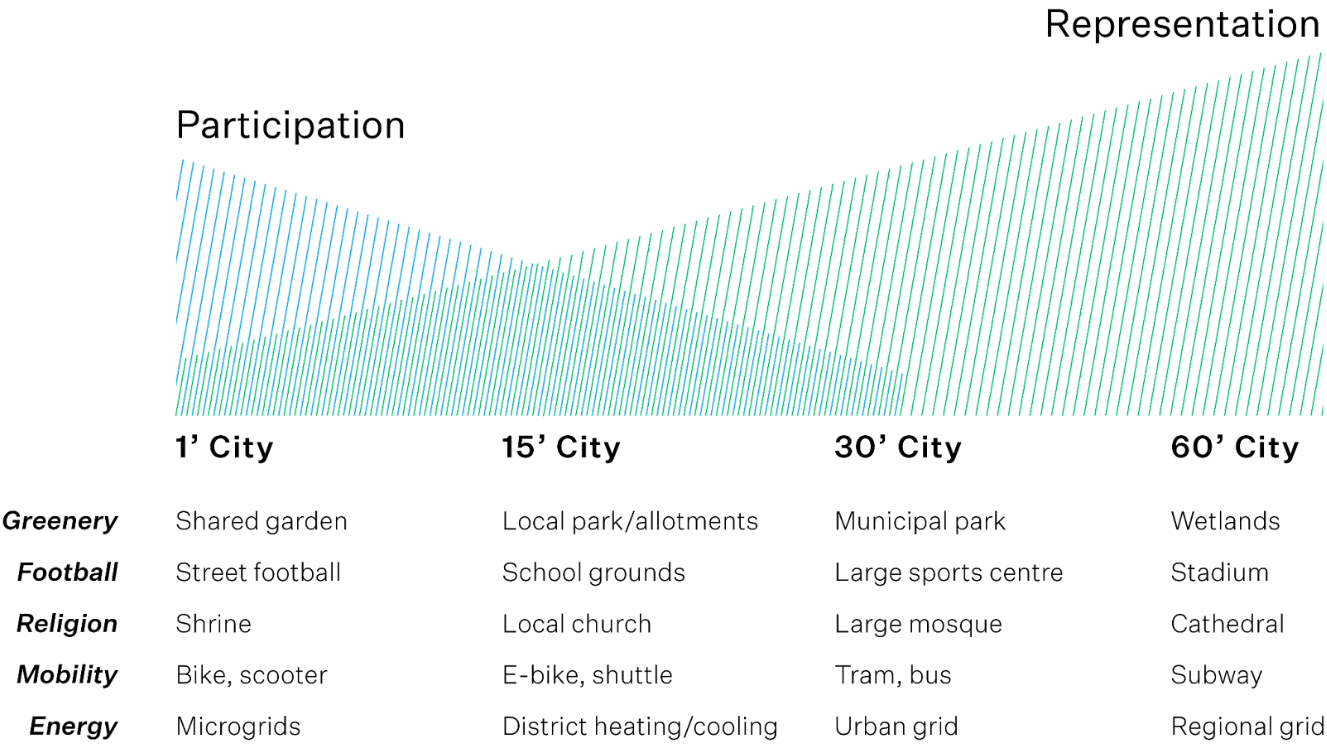}
\caption{\textbf{Schematic diagram illustrating the interplay between the intrinsic multiscale structure of cities and the full spectrum of agency from Participation, i.e., bottom-up approaches, to Representation, i.e., top-down approaches.}}
\label{fig:participation_scales}
\end{figure}
As discussed, in practice, the governance model derived from a proximity city-based policy framework for 15-minute cities largely organises relatively large urban scales — neighbourhoods, districts, zones — using municipal planning and service delivery processes. In contrast, an effective 1-minute city scale is necessarily more participative (its scale and diversity could only be realistically organised by allowing citizens to manage such spaces and services themselves). Equally, larger systems, again, such as city-wide infrastructure systems, maybe '60-minute cities' or more in terms of this notional elision of scale and time. Diagramming these systems across these different scales indicates a way of thinking and acting about participative decision-making varying accordingly. 

As noted earlier in our Swedish example, a local street could be organised, to some considerable extent, at the 1-minute City scale via participative cultures of decision-making, with residents and users to the fore. These spaces could potentially be malleable, adaptable, open and diverse environments, and our emerging data schema must capture qualitative aspects accordingly, beyond their mere existence. Other mobility systems and environments, such as a subway, are large, expensive to create and run, and thus purposely challenging to adapt. Therefore, they tend to be organised around representative decision-making cultures and might best be informed by professional expertise and deliberative political processes in \textit{consultation} with members of the public. The purpose of decision-making at this scale is to produce equitable, reliable and robust outcomes at a 60-minute city scale across the city rather than diverse, adaptable environments. While there are qualitative aspects to the data collection for these larger systems too—quality of service, for instance—and while they also touch down in neighbourhoods, as systems, they are essentially designed and operated at the city scale. There are highly complex decisions at both ends of this spectrum, not least as these decisions are nested, i.e., decisions taken at the 1-minute scale affect those at the 60-minute scale and vice versa. This point is easy to see with the design of mobility and energy systems. Still, as the diagram suggests, it may also apply to questions of culture, like football, music, or religion. 
\begin{figure}[h]
\centering
\includegraphics[width=0.8\columnwidth]{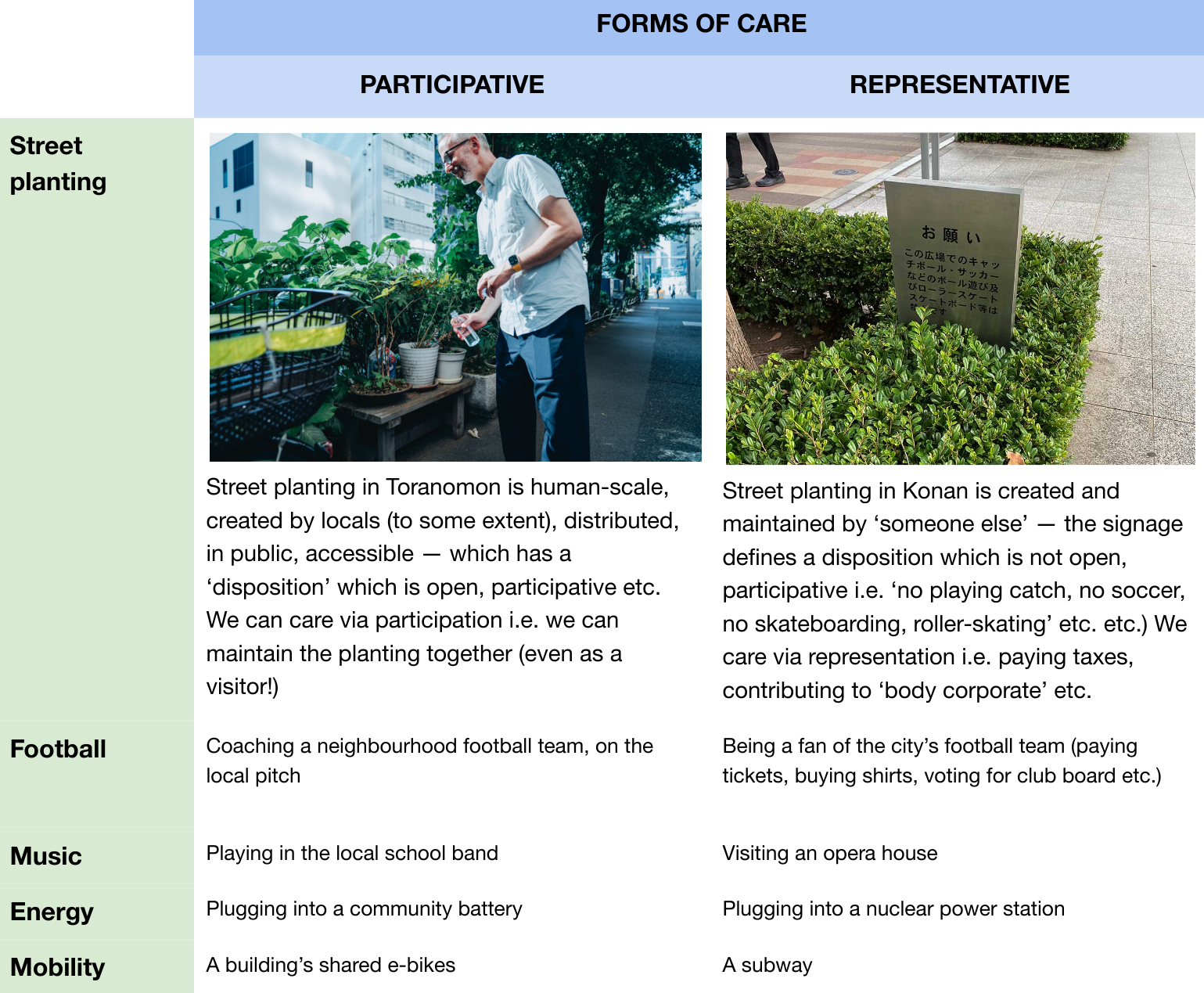}
\caption{\textbf{Different kinds of care.}}
\label{fig:care_forms}
\end{figure}
These two poles — 1-minute and 60-minute, participative and representative — denote different forms of decision-making and thus offer different affordances for, and qualities of, participation and care. One might care about the existence of a subway system, yet it is highly unlikely that citizens will be asked to maintain or manage it directly. Instead, they contribute to the system through representative forms of care, such as via taxation and democratic processes. On the other hand, a community garden, at street scale, can be maintained through directly participative and inclusive processes. It is deliberately not 'outsourced' to municipal government.

These patterns could exist across different systems, offering a different way of thinking about nested scales of decision-making and care.

\review{Scaling participatory approaches in large urban areas presents significant challenges, especially in maintaining the balance between representation and participation at different scales. However, successful models have demonstrated that scaling is possible with the right frameworks and support mechanisms. For instance, Porto Alegre’s participatory budgeting has empowered citizens to participate in budgetary decisions at a municipal level, scaling input from local communities to the entire city \cite{baiocchi2001participation,wampler2012participatory}. Similarly, Seoul’s citizen hackathons have used digital platforms to enable large-scale participation in shaping urban policies and projects \cite{smith_2018}. Platforms like Decidim, particularly in Barcelona, and vTaiwan, more nationally, have developed sophisticated constructive co-creation and deliberation methods which suggest possible futures for urban governance \cite{helbing2023democracy}.

The success of these initiatives is rooted in a combination of digital tools for broad engagement and on-the-ground partnerships with local organizations to mobilize and gather diverse inputs. Indonesia’s One Map Initiative further highlights how integrating community input into national land-use planning can bridge the gap between local and governmental perspectives, showing that participatory approaches can influence policy at the highest levels when adequately supported \cite{shahab2016indonesia}. Equally, in the context of disaster recovery and resilience planning, the Peta Bancana platform \footnote{\href{https://info.petabencana.id/about/}{https://info.petabencana.id/about/}}, also based in Indonesia, indicates how open-source platforms can produce highly meaningful real-time disaster maps using both crowd-sourced reporting via social media correlated against government agency validation. Furthermore, the continued successes of platforms such as iNaturalist \cite{matheson2014inaturalist} and OpenStreetMap \cite{OpenStreetMap} suggest that with clever design and sustained effort, some forms of citizen participation can even reach global scale.

Ensuring diverse demographic representation is critical in scaling these processes. Targeted outreach to underrepresented groups, the provision of accessible participation channels, and the training of local facilitators are key strategies to mitigate biases inherent in bottom-up approaches. San Francisco’s Groundplay parklet initiative exemplifies how small-scale, localized interventions can scale equitably by providing opportunities for citizens to design and manage public spaces, supported by both City and County of San Francisco. As with the city’s Shared Spaces program \footnote{\href{https://www.sf.gov/shared-spaces}{https://www.sf.gov/shared-spaces}}, this institutional ‘care’, combined with user-friendly guides and the parklet’s inherent possibility of decentralised interventions, can foster sustained engagement and can help ensure that all voices are heard. Vinnova’s ‘1-minute city’ initiative equally uses parklet-like models to garner citizen engagement at scale, providing a scalable, spreadable system for multiple Swedish municipalities, yet customised by citizens directly \cite{hill2022designing}.}

\subsection{Towards cities of opportunities}

\begin{figure}[ht]
\centering
\includegraphics[width=0.8\columnwidth]{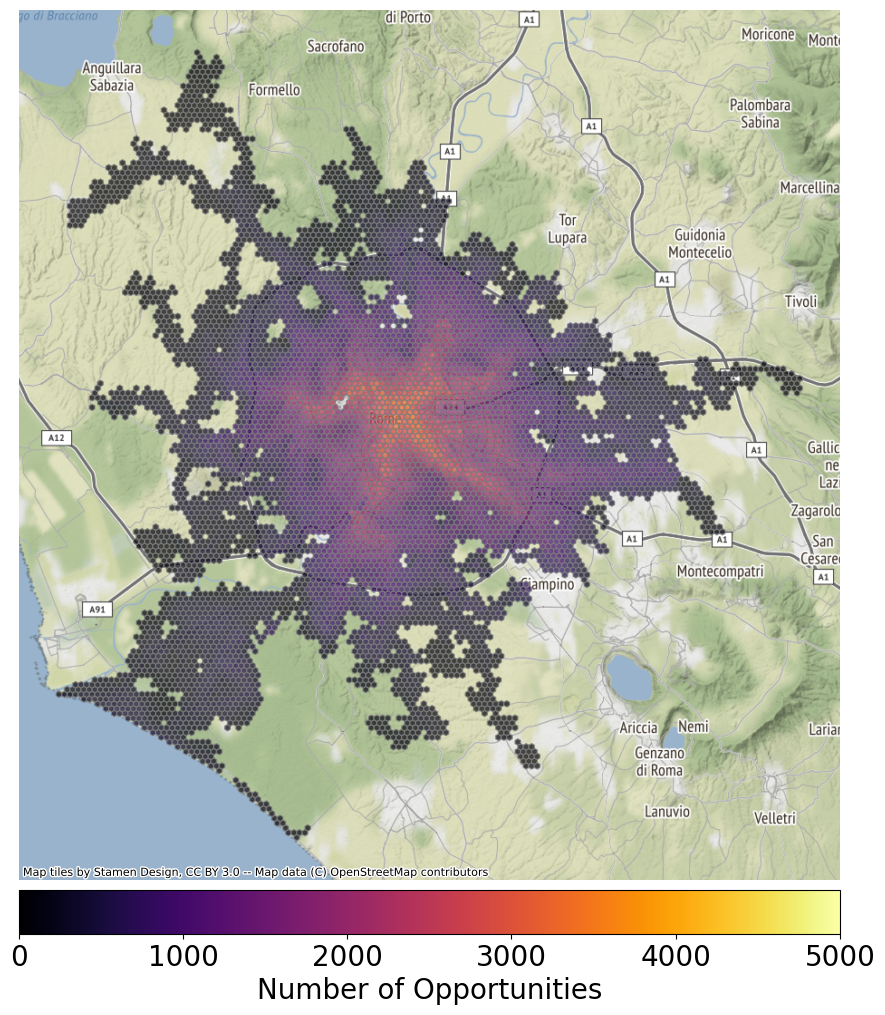}
\caption{\textbf{A visualisation of Rome as a \textit{city of opportunities}}. Colours encode the number of opportunities accessible from every location in the city. Accessibility here is computed as the number of amenities reachable by public transportation and walking, with a typical trip of a city dweller. The methodology is analogous to that introduced to measure the "sociality score" in~\cite{biazzo2019general}.}
\label{fig:city_of_opportunity}
\end{figure}

The importance of amenities and services closer, in the spectrum of Participation-Representation, to Representation cannot be neglected. Research showed that proximity access cannot be comprehensive: even with services close to their home, residents have been shown to travel further than strictly needed~\cite{abbiasov2022quantified}. This element represents one of the main advantages of cities: they can afford expensive services given the number of people they host. We are talking about large stadiums, monuments, theatres, large parks, big sporting venues, hospitals, and universities. These cannot be proximity services: such large amenities require a substantial user base to be economically sustainable. Large cities exist to support this whole ecosystem, and the recent trends of rapid urbanisation hint at the fact that this ecosystem is attractive to people. Therefore, we should encompass this aspect when modelling access and a healthy, sustainable lifestyle. The 15-minute city model does not take into account this long-range access, to the point that it is sometimes perceived as a way to restrict the mobility of people\footnote{\href{https://www.wired.com/story/15-minute-cities-conspiracy-climate-denier/}{https://www.wired.com/story/15-minute-cities-conspiracy-climate-denier/}}.

To address this critical point, we can extend the 15-minute city concept using a more extensive chrono-urbanism concept, the \textit{city of opportunities}. As well as quantifying the number of local opportunities and distance of proximity services of urban areas, we can investigate the needs of citizens by capturing the local \textit{and} distant opportunities that are reachable by sustainable transportation. Fig.~\ref{fig:city_of_opportunity} presents a quantitative representation of global access to opportunities in Rome, Italy. More precisely, the map in the figure shows a grid coloured according to the number of POIs accessible by every area in the city through possible combinations of public transportation and walking trips. The number of accessible opportunities is highly heterogeneous across the city, signalling that central regions allow access to almost all opportunities (including landmarking locations). At the same time, an efficient public transportation system greatly reduces the disparity: here, the metro lines of Rome boost the number of opportunities reachable from intermediate and outer areas. No proximity city could allow the healing of that highly unequal situation. However, efficient public transportation implementations can: the underlying idea is that citizens should be able to access all types of opportunities, whether close or far. Most types of opportunities should be within walking or biking distance. On the contrary, more representation-like amenities and services, such as prominent theatres, stadiums or landmarks, should be equally accessible through public transportation.

\section{Developing New Data-based Approaches}\label{sec:data}

Throughout this paper, we have advocated for a transition toward a richer form of urban analysis that can take into account a range of subjective, value-based metrics, moving away from the chrono-urbanism attitude that attempts to distil urban complexity into a single parameter, i.e., proximity. However, we must keep in mind that the proximity-based approach is \textit{necessarily} simplistic, conforming to the current availability of data. While proximity-based urban analysis only requires limited sets of data such as POI types/locations, information on road and public transit networks, etc., a more comprehensive value-based analysis will naturally require richer sets of urban data, many of which, as of now, cannot be sourced reliably and at scale. The same can be said about models and techniques for data analysis as well; investigations are needed to devise the specific ways in which richer, heterogeneous urban data can be integrated to enhance the proximity-based framework.

\review{In order to effectively gather and analyse qualitative data, urban planners can use a combination of methods such as surveys, focus groups, and digital platforms. Tools like sentiment analysis of social media, participatory GIS mapping, and machine learning models for pattern recognition can offer deeper insights into community needs and preferences. It is important to ensure diverse demographic participation through targeted outreach and inclusive design of data collection methods for comprehensive analysis. Particular care should be taken when relying on digital tools for data acquisition, for example implementing strict adherence to accessibility standards and incorporating best practices from algorithmic fairness research.}

In this section, we discuss how current quantitative approaches fail to grasp the complexity that creates well-being in a local community and suggest some innovative paths forward.

\subsection{Towards a more comprehensive quantitative approach}

So far, data-based approaches have largely tackled urban issues from a quantitative perspective, modelling the discernible and easily measurable aspects of cities such as population density, mobility systems, land use, air quality, noise, socioeconomic factors, the proximity of services, etc. These analyses are vital, and understanding such data is critical to discovering and delivering sustainable solutions and generative environments in cities~\cite{caldarelli2023role}. Nevertheless, more opaque or variable aspects of urban life, which may be less easily quantifiable, measurable, and thus less represented as data, are at least equally important. These aspects include community engagement, cultural identity, quality of life, perceptions of beauty, public participation, inclusion, identity, belonging, agency—and what Keller Easterling defines as disposition~\cite{easterling2014extrastatecraft}.

As Saskia Sassen has described, a city is a complex, incomplete system. These conditions enable cities to constantly remake themselves, with a resilience drawn from complexity and diversity that has given "cities their long lives across enormously diverse historical periods"~\cite{sassen_2017}. However, the particles that make up this urban system, i.e., citizens and other living elements, are complex agents, too. In this scenario, understanding the complexities and the laws that regulate the larger system is one (important) side, but understanding the laws that drive the behaviour of the particles that make up the city is another matter. Why would we create sustainable cities if not for the well-being of humanity and the world we live in? 

We encourage a new link between behavioural analyses and models of cities, which could incorporate individual well-being and societal benefits towards a comprehensive understanding of the city. Unifying the two approaches means finding the right balance between top-down regulations and bottom-up participation.

So far, the perspective of data-based approaches has mostly been on understanding and underscoring planning and design principles for infrastructures such as intersections, streets, squares, and the building blocks of neighbourhoods. We advocate for a change of focus: to introduce, study and work on data and models which take into consideration urban qualities, predicated as much on flows, systems, networks, cultures, interactions, and perceptions, to better advocate for collective vitality and well-being of both humans and more-than-human environmental perspectives. 

To model care and participation in urban society, modern science should, therefore, not only focus on ticking the items on the accessibility lists given by chrono-urbanism theories but also understand how to improve the urban experience of a diverse, complex, mixed-use neighbourhood. This is easier said than done: ticking categories in a basket of essential services or amenities complies with the classical rules of modern computers, turning them on and off bit by bit. However, data about lively communities, active participation, and the complex, more subtle aspects that make urban life enjoyable rather than acceptable are scarce. What kind of data would we need to design a vibrant and enjoyable neighbourhood? How could we measure the well-being of urban life, and is it even possible to grasp this aspect? What makes an open city?

\subsection{Finding new avenues for city data collection}

Static geolocalised data~\cite{OpenStreetMap} can be used to analyse spatial properties of urban settlements, but often lack detailed qualitative metadata. Even POIs containing reviews, scores, and so on are biased by the influence of the socio-cultural context in which they are immersed, the presence of tourism, and more local factors. Indeed, most analyses on proximity are currently run through this kind of data.

An important step forward has been utilising anonymised mobility data from single user~\cite{barbosa2018human}. This approach can shed light on the single users' patterns and preferences at a microscopic scale. However, even this kind of data presents some issues. First, the scale at which these analyses work is not so refined. Location-based data is usually reliable at a granular level but lacks an unbiased, statistically significant rigour at small scales. Furthermore, using anonymised data makes it impossible to understand the underrepresented categories of society, and the behavioural drivers of a single agent remain debatable.

Surveys still represent a unique tool that can be used to assess the public's perception regarding certain aspects. It is costly in time and resources to produce a comprehensive dataset that can extract significant insight at a microscopic level. Most currently available surveys ask essential questions at the aggregated level of a city, sometimes a district. However, we need to understand more microscopic patterns of care, perception, and personal experience to build enjoyable and lively communities. Surveys describing how places are perceived in the small scale, such as rating street view images in their safety and enjoyability, try to bridge this gap~\cite{wu2024integrating}.

Social network data can also help in understanding the sentiment of urban dwellers towards the surrounding environment and parts of the city~\cite{frias2012characterizing,rowe2023urban}. Similarly to mobility data, this kind of data is often limited in its applicability for microscopic analyses due to the lack of granularity, noise and statistical bias. For commercial social networks, the absence of guarantees for continued data availability (APIs can be modified or discontinued) often becomes an issue as well.

Mixing social networks and surveys, the \href{dedede.de/en}{\textit{dédédé}} platform allows users to submit geolocalised posts about POIs and other objects or phenomena in cities, each consisting of a short text description and (optionally) images, audio, and some metadata. Each post contains qualitative feedback on a specific POI or urban entity, to which the community can show agreements, disagreements, and other opinions through interactive mechanisms (e.g., comments) offered on the platform. If used extensively and targeted to urban interventions, this kind of dedicated, transparent platform can help policymakers, researchers, and citizens make sense of elements present in the city without making specific assumptions on what is good or not for residents. Such approaches try to satisfy the need for unbiased, bottom-up feedback on the public perception of public spaces, much along the features described in Tables~\ref{table:1} and ~\ref{table:2}. Too often, the public's preferences for urban organisation are misinterpreted, starting from assumptions that evolve over time. The supposed good practices for urban settlements consistently lag behind technological and societal changes, rarely catching up. Such bottom-up built datasets could help understand the current perception of spaces and help fix urban problems locally. However, encouraging citizens to participate through a dedicated app is not trivial. It requires dedication and a continuous flow of information directly to public decision-makers, which can enhance public engagement by demonstrating the benefits of participation. Furthermore, even dedicated platforms operated without commercial motives still often adopt user interface elements and ideas (such as ``likes'') from popular for-profit platforms, which do not necessarily align with the goal of unbiased data collection \cite{morozov2013solutionism}; building a successful social platform that can assist value-based urban analysis will likely require a thorough rethinking of interaction patterns, metaphors, etc.

\review{While acquiring a richer diversity of data may not initially appear advantageous, incorporating data about personal experiences and preferences could significantly enhance decision-making, both institutional and communal, by offering a more nuanced understanding of community needs. Policymakers, planners, business owners, and community organisers could utilize this citizen- and environment-generated data to better understand and define potential policies, services and activities. This opens up a gradient of cultures of decision-making, from participative to representative. In different ‘administrative traditions’, local governments such as municipalities have widely diverging capabilities and responsibilities \cite{peters2021administrative}. Some municipalities deliver most everyday services and infrastructures, with significant fiscal powers; others, not at all. Yet each might be able to maximize the impact of policies they are responsible for, in terms of impacting overall health, well-being and sustainability of people and place, by ensuring that their interventions are designed and distributed equitably and effectively according to both individual and collective needs and desires. Some of these new public capabilities are manifesting themselves in place-based wealth funds, enabling public or common good ‘value capture’, based on participative governance approaches \cite{mazzucato2022mission}. Better understanding and enhancing those approaches, by working with qualitative data, is surely key to this new public capability, just as it is for business owners and community organisers.}

\begin{figure}[h]
\centering
\includegraphics[width=0.49\columnwidth]{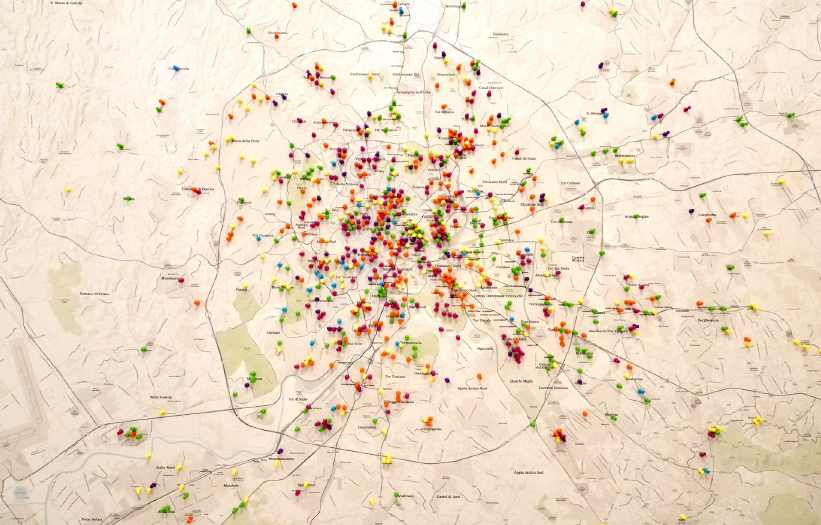}
\includegraphics[width=0.47\columnwidth]{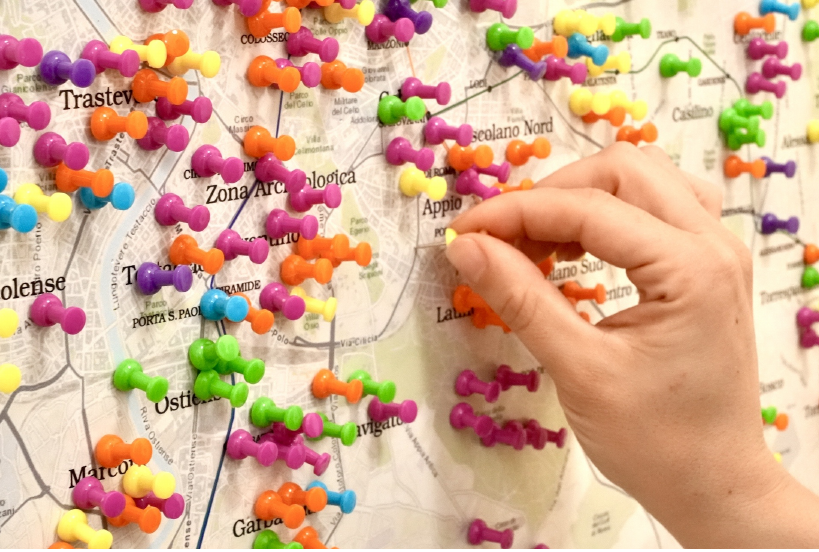}
\caption{\textbf{An "analogical" map of perceptions of Rome compiled through a bottom-up approach.}}
\label{fig:15minute-map}
\end{figure}

Despite difficulties as discussed above, data collection through public participation can nicely complement quantitative analyses based on objective data; here, we report a simple example that demonstrates this fact. Figure~\ref{fig:15minute-map} report the picture of an “analogic” map of perceptions of Rome compiled in a bottom-up way by volunteers in recent open events organised by Sony CSL. Volunteers were asked to pin down, on a physical map, locations in the city labelled with four colour codes corresponding to several categories: “It is nice”, “I feel good", “It sucks”, and “I need”. Although the perception map never had the presumption of becoming a scientific instrument, one can still draw some conclusions. If one compares this map with a representation of the 15-minute city of Rome, one immediately realises the richness added by individual contributions. For instance, focusing on the city centre, the map of perceptions features a great diversity beyond the single-minded category of accessibility time.  

\section{Conclusion}

The discourse surrounding proximity-based cities, particularly exemplified by the 15-minute city model, has illuminated the critical significance of accessibility and convenience in shaping urban environments. However, a nuanced examination reveals that a single-minded emphasis on proximity risks oversimplifying the complex dynamics of urban life. While the 15-minute city rightly champions the need for essential services and amenities within a short distance, it often falls short in accounting for the multifaceted dimensions of urban vibrancy, community identity, and quality of life.

Our analysis underscores the imperative of transitioning towards value-based cities, which embrace a broader spectrum of considerations beyond mere proximity. By incorporating factors such as diversity, quality, value, and social relationships into urban planning frameworks, cities can cultivate stronger local identities, foster resilience, and bolster democratic participation among residents.

\review{Moreover, we contend that achieving this transition necessitates a balanced approach that harmonizes both bottom-up and top-down interventions. While top-down strategies wield considerable influence in shaping urban landscapes, their overemphasis risks exacerbating existing inequalities and marginalizing grassroots initiatives. Conversely, grassroots efforts, while invaluable in fostering community cohesion and innovation, may encounter barriers to scalability and resource mobilisation without complementary support from governmental and institutional structures. Also, bottom-up approaches are not without challenges, as they can introduce biases if underrepresented groups are not properly included. To mitigate these biases, it is crucial to implement inclusive engagement strategies, such as targeted outreach, accessible participation channels, and training local facilitators, including the creation of city government participative design teams and shared public platforms for community organisations. Any technical solutions used to facilitate citizen participation (such as digital platforms, machine learning algorithms, etc.) must actively counter biases by being recognised as core public infrastructure and designed, delivered and operated as such. By synergizing these approaches, and building these new capabilities and infrastructures, cities can harness the strengths of both grassroots activism and centralised governance to maximize resilience, health, and sustainability, thereby mitigating disparities and fostering inclusive growth.}

Looking ahead, it is incumbent upon urban scholars and practitioners to chart new research pathways that contribute to the evolution of more comprehensive urban paradigms. This objective entails the development of participatory methodologies that empower marginalised communities to actively engage in decision-making processes, thereby democratising urban planning and governance. The last consideration links two elements often disconnected in the work of scholars: planning the best solutions for urban systems and designing new processes for effective decision-making at all possible scales, in terms of areas covered and population concerned. This last link was indeed the key topic emphasised in this Special Issue and summarised by the title of the workshop held at the Complexity Science Hub in Vienna in September 2023: "Co-Creating the Future: Participatory Cities and Digital Governance". 

Additionally, embracing the complexity of urban systems and their interdependencies across various domains—from energy and food security to cultural vitality and public health—will be instrumental in designing holistic and adaptive urban solutions that cater to diverse needs and aspirations.

In conclusion, while the 15-minute city offers valuable insights into the importance of proximity in urban planning, cities must go beyond mere proximity to truly thrive. By embracing a value-based approach that prioritizes diversity, quality, and community engagement, cities can unlock their full potential as vibrant, equitable, and sustainable places for all residents.

\subsubsection*{Acknowledgements}
VL wishes to acknowledge the organisers and participants of the workshop "Co-Creating the Future: Participatory Cities and Digital Governance", held at the Complexity Science Hub Vienna in September 2023, for all the valuable insights received. MB, VL and HPMM warmly thank Milena Di Canio for her invaluable contributions leading to the realisation of the perception map of Rome.

\bibliographystyle{unsrt}

\end{document}